\documentclass[twocolumns]{aa} %
\usepackage{graphicx}
\usepackage{epstopdf}
\usepackage{deluxetable}
\usepackage{times,epsfig} 
\usepackage{mathrsfs}  
\usepackage{natbib}
\usepackage{amssymb}
\bibpunct{(}{)}{;}{a}{}{,} 
\usepackage[english]{babel}
\usepackage{longtable}
\usepackage[english]{babel}
\usepackage{url}
\usepackage[]{hyperref}

\authorrunning{Taddia et al.}
\titlerunning{iPTF15dtg: Late-time luminosity excess in a Type Ic supernova}

\begin{document}

\title{The luminous late-time emission of the type-Ic supernova iPTF15dtg - evidence for powering from a magnetar?}

\author{F. Taddia\inst{1} 
\and J. Sollerman\inst{1}
\and C. Fremling\inst{2}
\and E. Karamehmetoglu\inst{1}
\and C. Barbarino\inst{1}
\and R. Lunnan\inst{1}
\and S. West\inst{1}
\and A. Gal-Yam\inst{3}}

\institute{The Oskar Klein Centre, Department of Astronomy, Stockholm University, AlbaNova, 10691 Stockholm, Sweden.\\ \email{francesco.taddia@astro.su.se}
  \and Division of Physics, Math and Astronomy, California Institute of Technology, 1200 East California Boulevard, Pasadena, CA 91125, USA ; Cahill Center for Astrophysics, California Institute of Technology, Pasadena, CA 91125, USA
\and Benoziyo Center for Astrophysics, Weizmann Institute of Science, Rehovot 76100, Israel}

\date{Received; accepted}

\abstract
{The transient iPTF15dtg is a type-Ic supernova (SN) showing a broad light curve around maximum light, consistent with massive ejecta if we assume a radioactive-powering scenario.}
{We aim to study the late-time light curve of iPTF15dtg, which turned out to be extraordinarily luminous for a stripped-envelope (SE) SN, and investigate possible powering mechanisms.}
{We compare the observed light curves to those of other
  SE~SNe and also to models for the $^{56}$Co decay. 
  We analyze and compare the spectra to nebular spectra of other
SE~SNe. We build a bolometric light curve and fit it with 
different models, including powering by radioactivity, magnetar powering, and a combination of the two.}
{Between 150 and 750~d post-explosion, the  luminosity of iPTF15dtg declined by merely two magnitudes instead of the six magnitudes expected from $^{56}$Co decay. This is the first
spectroscopically regular SE~SN found to show this behavior. 
The model with both
radioactivity and magnetar powering provides the best fit to the light curve and appears to be the most realistic powering mechanism. An alternative mechanism might be circumstellar-medium (CSM) interaction.
However, the spectra of iPTF15dtg are very
similar to those of other SE~SNe, and do not show signs of strong CSM
interaction.
}
{The object iPTF15dtg is the first spectroscopically regular SE~SN whose light curve displays such clear signs of a magnetar contributing to its late-time 
powering. Given this result, 
the mass of the ejecta needs to be revised to a lower 
value, and therefore the progenitor mass could be 
significantly lower than the previously estimated $>$35 $M_{\odot}$.}


\maketitle

\section{Introduction}

Core-collapse (CC) stripped-envelope (SE) supernovae (SNe), such as
Type IIb, Ib and Ic, are believed to be powered by radioactive decay
of $^{56}$Ni  and its decay product $^{56}$Co. When radioactive
elements such as $^{56}$Co are the source of the SN emission, their
characteristic decay times should govern the evolution of the SN
luminosity, especially at late epochs when the ejecta are mostly transparent. 
In fact, the late-time observations of SE~SNe
have so far revealed that the luminosity is declining at the rate of
$^{56}$Co decay or in most cases faster than that, due to increasing
gamma-ray leakage from the ejecta \citep[e.g.,][]{wheeler15}. Hitherto we have
not observed a spectroscopically regular SE~SN whose luminosity declines slower than the
$^{56}$Co decay (0.0098 mag~day$^{-1}$), which agrees well with the
overall idea that CC SE~SNe are radioactively powered. 

\begin{figure*}
\includegraphics[width=18cm]{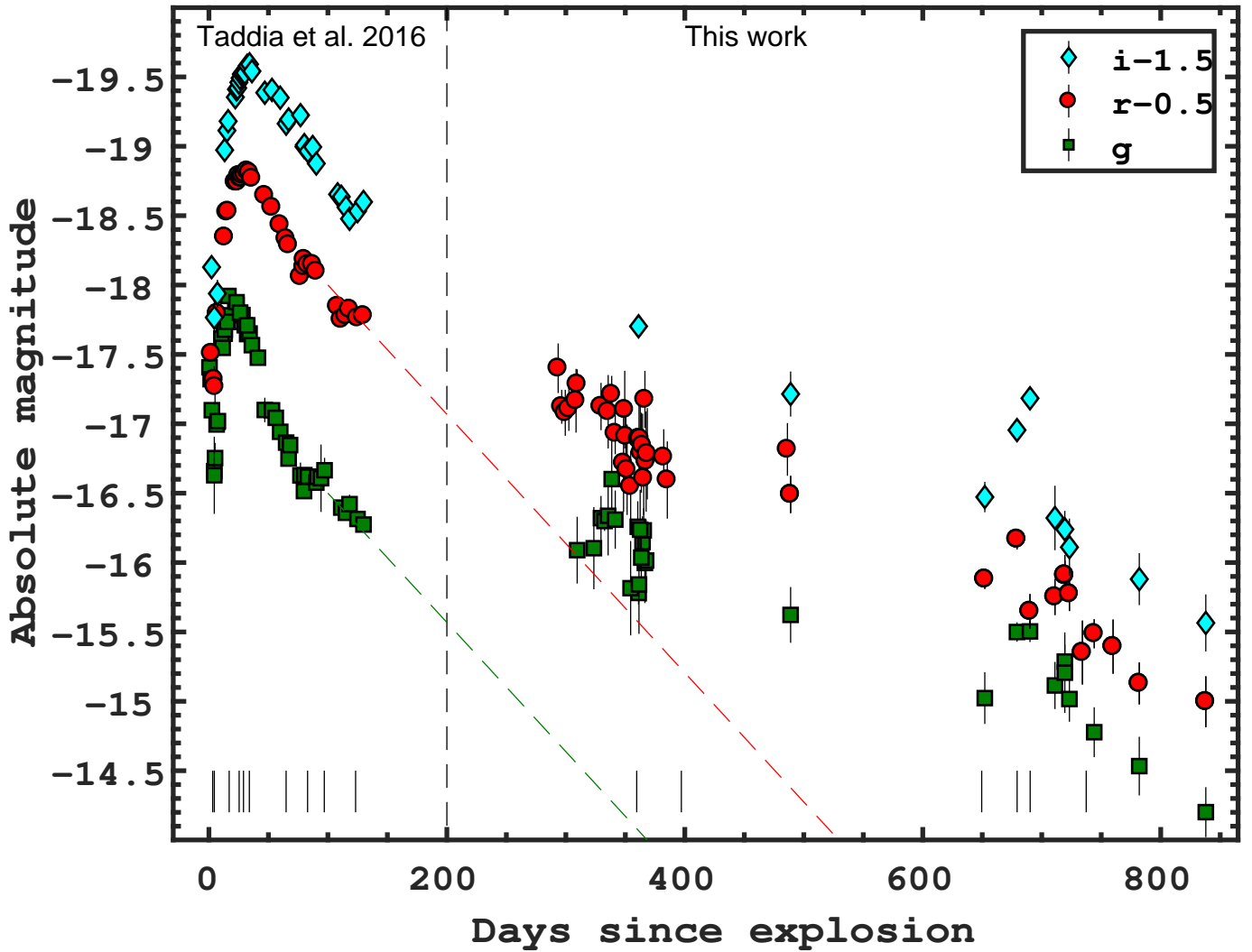}
\caption{\label{lateLC} Absolute-magnitude light curves of iPTF15dtg in the $gri$ filters in the observer frame.
Each light curve was shifted by the amount indicated in the legend for clarity. The spectral epochs are reported as vertical black segments at the bottom. The $^{56}$Co decay rate is shown as dashed lines and scaled to both early $g$ and $r$ band photometry.}
\end{figure*}

The late-time luminosity of a CC~SN could decline more slowly than the
$^{56}$Co decay rate if a different or additional powering
source is at play. This could for example be circumstellar interaction or the
energy input from a magnetar. Some type-IIn SNe exhibit long-lasting emission at high luminosity, because their ejecta interact with their surrounding medium and convert the kinetic energy into radiation \citep[see e.g.,][for the case of SNe~2005ip and 2006jd]{stritzinger12}. 
Such interaction also produces features in the spectra (narrow
emission lines) that are easy to identify. An example of a CSM-interacting SE~SN is SN~2010mb \citep{benami14}, whose spectrum shows a remarkable blue color and whose light curve exhibits a slow decline.  
Superluminous SNe (SLSNe) are believed to be magnetar-powered, as their late phases are also particularly bright and difficult to reconcile with $^{56}$Co powering \citep{inserra13}.

\citet{taddia16} (hereafter T16) presented and analyzed the first 150 days of SN iPTF15dtg. This type-Ic SN exhibits a particularly broad light curve, similar to that of SN~2011bm (\citealp{valenti12}). 
Its broadness suggests massive ejecta in a scenario where radioactive decay powers the SN. 
The very early epochs of iPTF15dtg also revealed a first peak on the
timescale of a few days, possibly powered by the post-shock breakout
cooling tail from an extended envelope. Signs of such early emission were also found in the early data of SN~2011bm (T16).
The early spectra of iPTF15dtg are similar to those of normal type-Ic
SNe, and show no clear signs of interaction with CSM.

After about five months of follow up, iPTF15dtg went behind the Sun. This
is when the initial monitoring was terminated and the early results were published in T16. 
After approximately 300 days post-explosion, we recovered the SN with the P48
telescope at Palomar $-$ to our great surprise. Where we had 
expected the SN to decline at least as
fast as the decay time of $^{56}$Co, it appeared much brighter. 
We therefore started our late-time monitoring and kept following it until $\sim$840 days, as it declined very
slowly. iPTF15dtg is indeed 
declining slower than
radioactive decay, revealing a new powering mechanism for its late
epochs. 
We argue below that this mechanism could be the spin-down of a magnetar, 
combined with the classic radioactive emission. 

In this paper, we present the late-time data acquisition and reduction for iPTF15dtg in Sect.~\ref{sec:dataac}. We present and analyze its late-time light curves and the spectra in Sects. ~\ref{sec:lc} and~\ref{sec:spec}, respectively. 
We build (Sect.~\ref{sec:bolo})
and model the bolometric light curve in Sect.~\ref{sec:model}. 
We then discuss the powering mechanism and the
implications for the progenitor star in Sect.~\ref{sec:discussion} and
finally give our conclusions in Sect.~\ref{sec:concl}. 

\section{Data acquisition and reduction}
\label{sec:dataac}

\begin{figure}
\includegraphics[width=9cm]{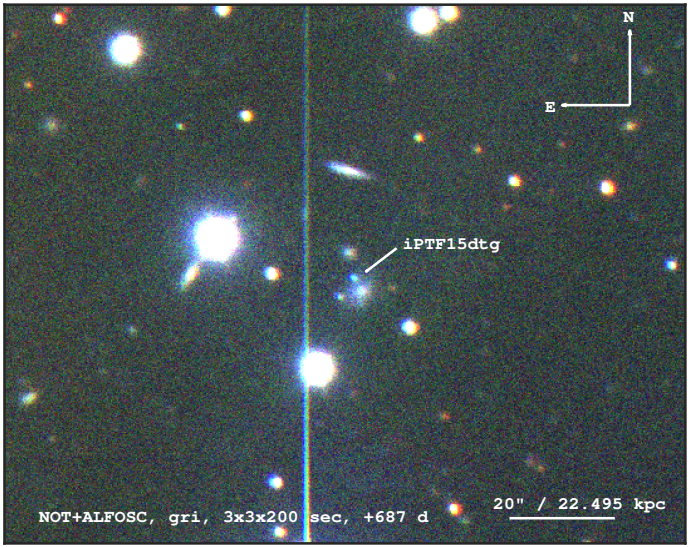}
\caption{\label{FClate}iPTF15dtg in a color-combined image from the NOT equipped with ALFOSC taken 687~d post-explosion. The SN is still remarkably bright compared to its host. We report the scale and the orientation of the image in the right-hand corners. A saturated star outside of the field-of-view produced the vertical feature in the middle of the frame.}
\end{figure}

\begin{figure}
\includegraphics[width=9cm]{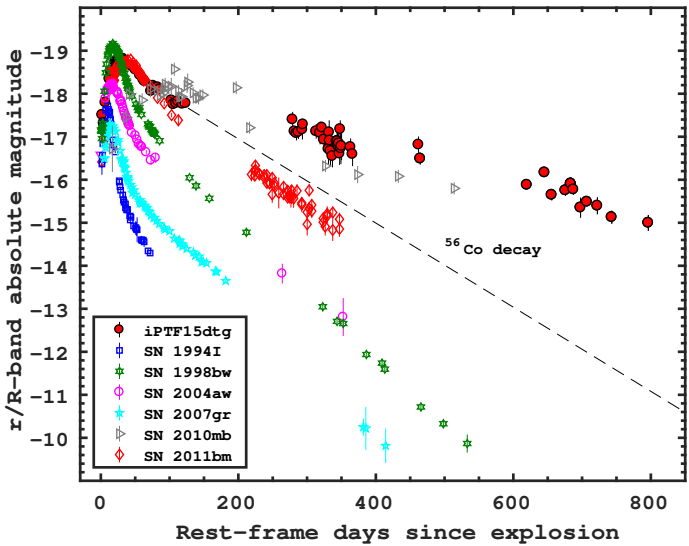}
\caption{\label{complateLC}The $r/R$-band absolute-magnitude light curves of a few type-Ic SNe from the literature, compared to that of iPTF15dtg, which is declining much more slowly. The references for these light curves are listed in Fig. 3 of T16. Supernova~2010mb data are from \citet{benami14}. The late-time $^{56}$Ni + $^{56}$Co luminosity expected from the amount of $^{56}$Ni estimated from the peak of iPTF15dtg is marked by a dashed black line.}
\end{figure}

\begin{figure}
\includegraphics[width=9cm]{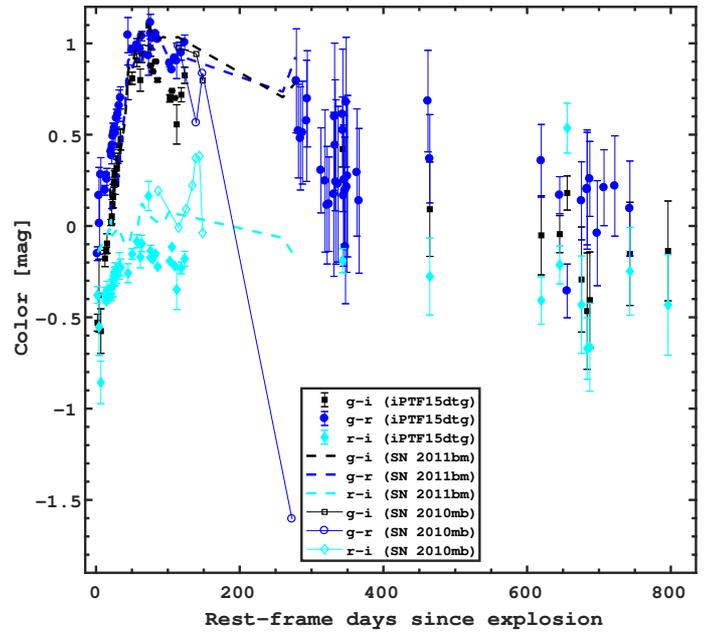}
\caption{\label{color}Color comparison between iPTF15dtg, SN~2011bm \citep{valenti12}, and SN~2010mb \citep{benami14}. At late epochs ($\sim$270~d) SN~2010mb exhibits a much bluer $g-r$ color than the other two SNe.}
\end{figure}

Our late-time data for iPTF15dtg are from a number of facilities.  Photometry in $g$ and $r$ band was obtained with the P48 telescope equipped with the 96 Mpixel mosaic camera CFH12K \citep{rahmer08}. The Nordic Optical Telescope (NOT) with ALFOSC (Andalucia Faint Object Spectrograph and Camera), the Telescopio Nazionale Galileo (TNG) with DOLORES (Device Optimized for the LOw RESolution), and the Palomar 200-inch Hale Telescope (P200) provided further $gri$ band photometry. A single epoch of $r$-band photometry was obtained with the Palomar 60" telescope (P60; \citealp{cenko06}) equipped with GRBcam. 
P48 photometry was reduced with the Palomar
Transient Factory Image Differencing and Extraction (PTFIDE)
pipeline which performs template subtraction and PSF photometry \citep{masci17}. 
The images from the NOT, TNG, P200, and P60 were reduced with the FPipe pipeline \citep{fremling16}. 

Standard stars from the PANSTARRS catalog\footnote{\href{http://archive.stsci.edu/panstarrs/search.php}{http://archive.stsci.edu/panstarrs/search.php}} were used as reference stars to calibrate the photometry. 

From the PANSTARRS archive we also obtained the deep images used to template-subtract our SN frames. The final light curves
are presented after combining the magnitudes obtained from the same
night. A log of the photometry is provided in Table~\ref{tab:phot}.

Late-time spectra are from Keck I equipped with LRIS (Low Resolution Imaging Spectrometer, \citealp{oke95}) and Keck II
equipped with DEIMOS, as well as from the TNG equipped with DOLORES
(grism LR-B and LR-R, slit 1$\farcs$0) and NOT equipped with ALFOSC (grism
\#4, slit 1$\farcs$0). 
All the spectra were reduced in a standard manner with dedicated
pipelines. The reduction includes bias correction and flat-fielding,
one-dimensional (1D) spectral extraction, wavelength calibration through a comparison
with an arc-lamp, and flux calibration through a sensitivity function
derived from spectrophotometric standard stars. 
Absolute flux calibration was obtained by computing synthetic photometry
in the $r$ band from the spectra, by comparing the obtained photometry to
the $r$-band light curves and, when needed, by multiplying the spectra for the flux ratio correspondent to the difference in magnitude. A log of the spectroscopy is provided in Table~\ref{tab:spectra}.

\section{Late-time light curves}
\label{sec:lc}

\begin{figure*}
\includegraphics[width=18cm]{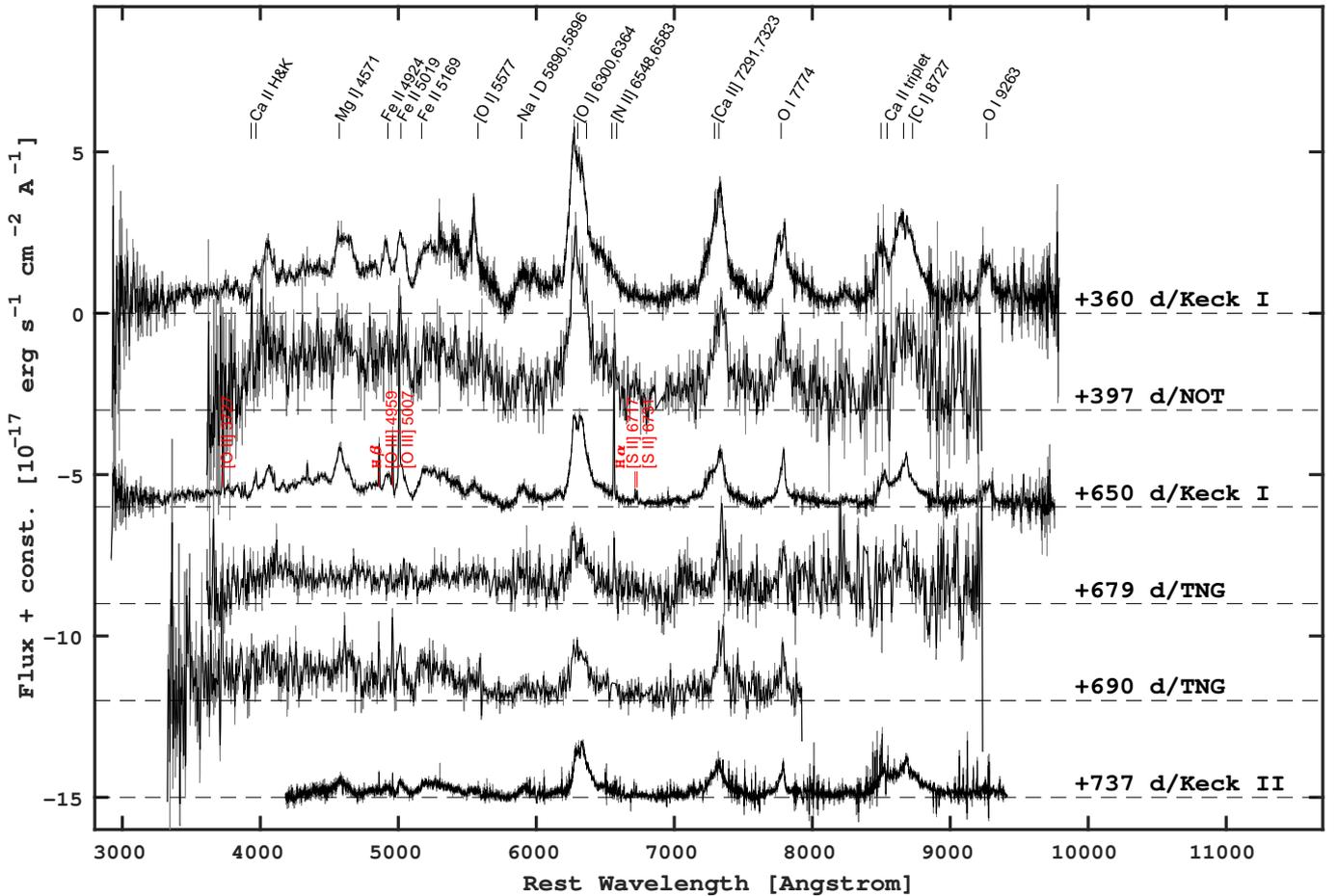}
\caption{\label{latespec}Late-time spectral sequence of iPTF15dtg. All the spectra are absolute-flux calibrated and shown in the rest frame. The zero-flux level for each spectrum is marked by a horizontal dashed line, as well as its phase and telescope. The main SN features are labeled in black at the top, those related to the host-galaxy emission, well visible in the third spectrum, are labeled in red.}
\end{figure*}

 \begin{figure}
\includegraphics[width=9cm]{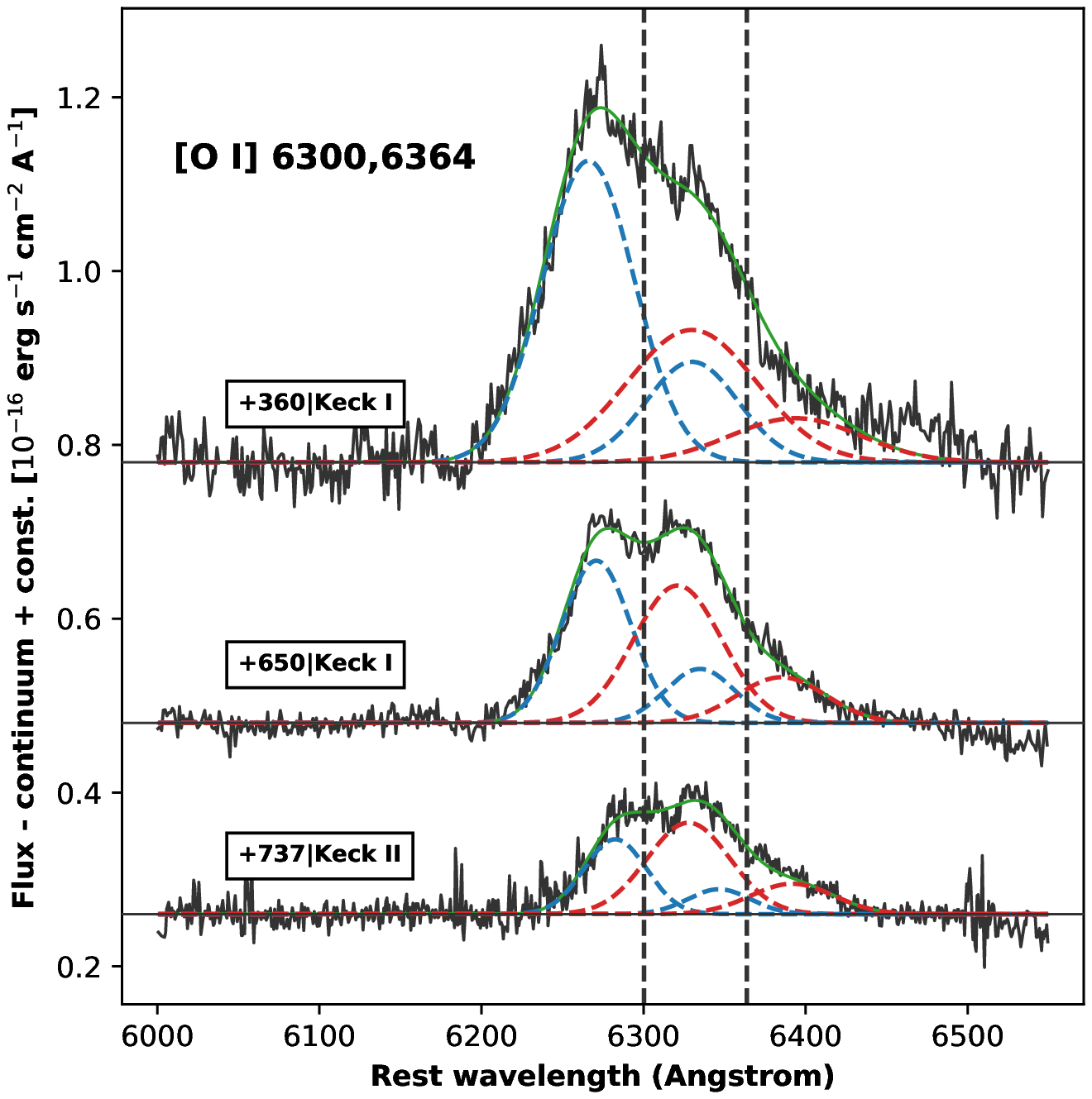}
\caption{\label{oi} Continuum-subtracted [\ion{O}{i}]~$\lambda$6300,6364 line profile fit. Two Gaussians (a blueshifted and a redshifted component) for each of the two lines (6300 and 6364~\AA) were used. The normalization of the two lines was fixed so that the 6300~\AA~line has three times the flux of the 6364~\AA~line. The centroids of the corresponding components of the two lines were also fixed to be 64~\AA~apart.}
\end{figure} 

The early-time ($<150$ d) light curves of iPTF15dtg were presented in T16. Later, P48 epochs in $g$ and $r$ bands cover 
$\sim300-400$ days since explosion. A single $i$-band detection from the NOT was obtained in that same period. 
An epoch of 
$gri$ band photometry from the NOT was also obtained around 500~d, together with a single P60 $r$-band point. Telescopio
Nazionale Galileo, NOT, and P200 $gri$ data cover the epochs between 650 and 840 d. In Fig.~\ref{lateLC} we plot the absolute-magnitude light curves obtained using the apparent magnitudes given in Table~\ref{tab:phot}, together with the explosion epoch, the distance modulus, and the extinction from T16. 

As mentioned above, the late time ($>300$ d) photometry of iPTF15dtg was surprisingly bright. 
A SE~SN is expected to
decline at least as fast as the $^{56}$Co decay \citep[see
  e.g.,][]{wheeler15}, which in Fig.~\ref{lateLC} is marked by dashed
lines for the $g$ (green) and $r$ (red) bands.
Assuming that the SN had settled on the linear
decay just before disappearing
behind the Sun at 150~d, we recovered it around
300 to 400 d, 1.4 mag brighter than expected in
the $r$ band. 

The brightness of iPTF15dtg continued to decline extremely slowly between 500 and 650~d. A relatively short (3$\times$600~s) set of $gri$ images of
iPTF15dtg taken with the NOT at 687~d, which is shown in
Fig.~\ref{FClate}, still show a bright SN as compared to its host
galaxy. This is unique for SE~SNe, which are typically extremely hard to
follow at such late epochs unless they are very nearby (iPTF15dtg exploded at 232~Mpc, T16). The decline in the $r$ band between 150 and 750 days
is only $\sim$2~mag, instead of the 6~mag expected from the $^{56}$Co decay. 

The anomalous behavior of the light curves of iPTF15dtg at late times is even more
evident when compared to the late-time light curves of other SE SNe, as shown in
Fig.~\ref{complateLC}. The SN that most resembles iPTF15dtg at early
epochs, that is, SN~2011bm, has a similar broad light curve
peak but declines at the  $^{56}$Co decay rate. This is as expected
from a SN with a massive envelope which will trap all the
gamma-rays. Other SE~SNe with narrower peaks decline even faster, as
their ejecta are unable to trap all the gamma rays. The SN iPTF15dtg is
instead well above the late-time luminosity expected if the same amount
of $^{56}$Ni (and its decay product $^{56}$Co) were to power both the peak and the late emission. 
The
decay luminosity 
from the amount of radioactive material necessary to power its peak, as derived in T16, is
marked by a dashed black line in Fig.~\ref{complateLC}. 
The only other SE~SN that is above the $^{56}$Co decay luminosity in this 
figure is SN~2010mb.  
However, that SN also
had a very peculiar early light curve. It had a more than 100 days
long rise, a sort of plateau around peak lasting another 100 days, and
signs of strong CSM interaction in the spectrum (see
Sect.~\ref{sec:spec}). In particular, SN~2010mb showed a very blue color of the
spectrum, seen also in the late time light curves \citep[][see also Fig.~\ref{color}]{benami14}. 
Therefore SN~2010mb cannot be considered a regular SE~SN. Another SE~SN declining slower than $^{56}$Co decay at late epochs is SN~2014C, whose $R$ band dropped by 2.16 mag between 139 and 509 d \citep[][ iPTF15dtg dropped by 1.29 mag in the same time range]{margutti17}. However, the late-time emission of SN~2014C is characterized by strong CSM interaction revealed by a prominent H$\alpha$ emission in the spectrum \citep{mili15}.

In Fig.~\ref{color} we compare the colors of iPTF15dtg to those of type-Ic SNe~2011bm and~2010mb. The late-time colors of SN~2011bm are only slightly redder than those of iPTF15dtg. The SN~2010mb shows a much bluer $g-r$ color than iPTF15dtg at 270~d. 

\section{Late-time spectra}
\label{sec:spec}

Whereas the late-time light curves are unique, the
late-time spectra of iPTF15dtg are very similar to those 
observed in regular SE~SNe \citep[e.g.,][]{taubenberger09,jerkstrand15}. 
 High-quality spectra from Keck I (LRIS) and Keck II (DEIMOS) 
 were obtained at 360, 650, and 737~d after explosion. Additional spectra from NOT (ALFOSC) and TNG (DOLORES) were obtained between these epochs. All the spectra will be released via WISeREP\footnote{\href{https://wiserep.weizmann.ac.il}{https://wiserep.weizmann.ac.il}} \citep{yaron12}. 
We show in Fig.~\ref{latespec} the absolute-flux calibrated 
spectra in the rest-frame. In the figure we also identify the 
main emission lines. The [\ion{O}{i}]~$\lambda$6300 line is the most
prominent feature, together with the [\ion{Ca}{ii}]~$\lambda\lambda$7291,7323 and the
\ion{Ca}{II} triplet. The oxygen lines from [\ion{O}{i}]~$\lambda$5577, \ion{O}{i}~$\lambda$7774 and  
\ion{O}{i}~$\lambda$9263 are also present. In the bluer part of the spectra we observe the presence of \ion{Mg}{i}]~$\lambda$4571 and \ion{Fe}{ii} lines. Narrow emission lines from the host galaxy are present in some of the spectra (already observed at early epochs, T16), when the slit was oriented in a way that includes a nearby \ion{H}{ii} region. These lines are certainly from the host and not due to CSM interaction, as we observe [\ion{O}{ii}]$~\lambda$3727, which is characteristic of the spectra from \ion{H}{ii} regions.  We removed the host-galaxy
narrow lines from the SN spectra, except from the third spectrum where they are labeled in red.

The [\ion{O}{i}]~$\lambda$6300 line shows a double peak after 650~d. We were able to fit the [\ion{O}{i}]~$\lambda$6300, 6364 feature (see Fig.~\ref{oi}) with two Gaussian components for each line, fixing the ratio between the 6364~\AA\ and the 6300~\AA\ component to 1:3 (suitable for the optically thin limit) and keeping their central wavelengths at constant distance as well as keeping their respective FWHM equal. The continuum is subtracted by fitting a first-order polynomial. The physical origin of these two Gaussian components might be explained by two lobes of oxygen-rich material belonging to the ejecta, one moving in the direction of the observer while the other one moves in the opposite direction. This would result in a blueshifted and a redshifted component.

Additionally, the [\ion{O}{i}]~$\lambda$5577 line is strong in the first spectrum and fades until it almost disappears in the following spectra. Since this line is more prominent at high densities, its disappearance supports the assumption of optically thin emission used to fit the [\ion{O}{i}]~$\lambda$6300, 6364 feature in the very late spectra (see Fig.~\ref{oi}). 

The \ion{Mg}{i}]~$\lambda$4571 line shows a boxy profile in the first spectrum, which later disappears. The \ion{O}{i}~$\lambda$7774 line becomes narrower with time, similar to the various \ion{Ca}{II} lines in the spectra. 

\begin{figure*}
\includegraphics[width=18cm]{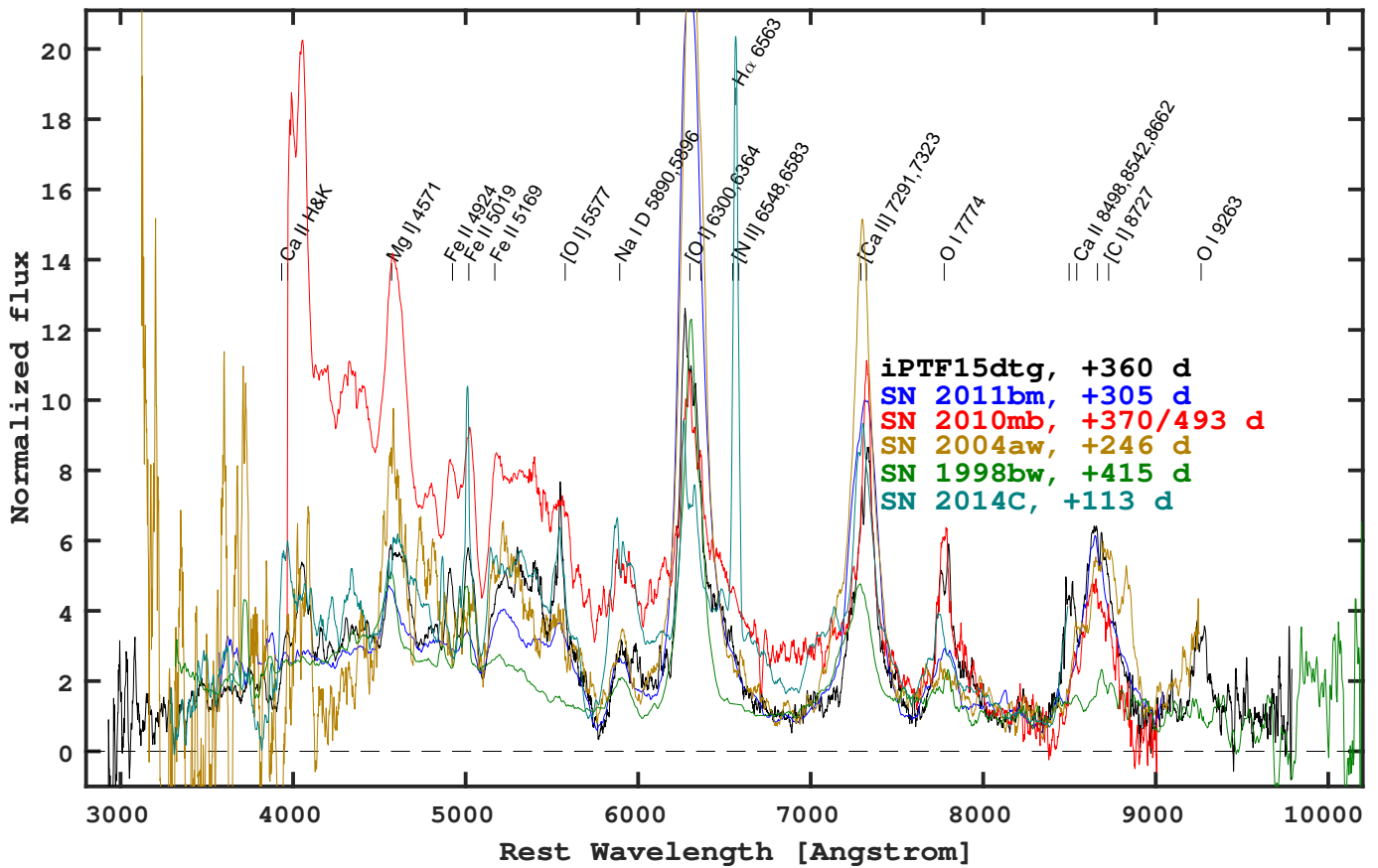}
\caption{\label{complatespec}Normalized (divided by the median flux between 8000 and 8400~\AA) spectra of iPTF15dtg (black), SN~2011bm (blue), SN~2010mb (red), SN~1998bw (green), SN~2004aw (orange), and SN~2014C (blue-green) at nebular epochs. All spectra were corrected for extinction. The main features are marked as in Fig.~\ref{latespec}. The phases are reported for each SN. Supernova~2010mb is clearly different as its continuum is remarkably bluer than for the other SNe. Supernova~2014C shows strong H$\alpha$ emission.}
\end{figure*}

When we compare the spectrum of iPTF15dtg with that of SN~2011bm around 300~d in Fig.~\ref{complatespec}, the two spectra are indeed very similar. Among the small differences, the [\ion{O}{i}]~$\lambda$5577 line is stronger (as compared to the continuum) in iPTF15dtg, and its \ion{Mg}{i}]~$\lambda$4571 line has a boxy profile. Around 
$4000-5500$~\AA\ our SN is marginally bluer than SN~2011bm. 
Furthermore, the [\ion{O}{i}]~$\lambda$6300 line is relatively (compared to the continuum) stronger in SN~2011bm. In Fig.~\ref{complatespec} we also compare the spectrum of iPTF15dtg at 360~d with late-time spectra of SN~Ic 2004aw \citep{taubenberger06} and SN~Ic-BL 1998bw \citep{patat01}. These spectra are also similar to that of iPTF15dtg, which is only slightly bluer. We also compare the spectrum of iPTF15dtg with that of SN~2014C from \citet{mili15}, in order to show the strong H$\alpha$ emission of that SN, indicative of CSM interaction. Supernova~2014C does not have a spectrum as blue as that of SN~2010mb. We also notice that the late-time \ion{O}{i}~7774 emission of iPTF15dtg resembles that of SLSN-I 2015bn \citep{nicholl15}, which is suggested to be powered by a central engine such as a magnetar. The late-time spectra of iPTF15dtg also resemble those of SN~Ib 2012au at similar phases \citep[][their Fig. 3]{mili18}. Interestingly, the  recent spectra of SN~2012au at six years after discovery reveal strong forbidden oxygen emission consistent with a pulsar wind nebula \citep{mili18}.

The spectra of iPTF15dtg do not reveal any clear sign of strong circumstellar-medium (CSM) interaction. Such an interaction could have explained the large excess in luminosity
described in Sect.~\ref{sec:lc}. The few SE SNe 
that are believed to be powered by CSM interaction, such as 
type Ibn
or the peculiar type-Ic SN~2010mb \citep{benami14}, do show
either strong (narrow) He lines in emission or a conspicuous blue continuum on top of the nebular spectrum. We show the 
spectrum of SN~2010mb at +300/400 d in Fig.~\ref{complatespec} (in red),
where the continuum displays a blue component,
which was interpreted as a sign of CSM interaction in \citet{benami14}. This is neither 
present in iPTF15dtg nor in SN~2011bm (or SN~1998bw and SN~2004aw). T16 reported the spectra of iPTF15dtg as being similar to those of SN~2011bm also at epochs earlier than 150~d. In short, in terms of 
spectroscopy iPTF15dtg simply resembles a spectroscopically regular SE~SN. 

\section{Late-time bolometric light curve}
\label{sec:bolo}

To address the nature of the powering mechanism of iPTF15dtg at late
epochs, we construct a bolometric light 
curve. The early bolometric light curve 
is adopted from T16, where it was 
computed using $g-r$ colors and the 
bolometric corrections by \citet{lyman14}. 
At later 
epochs, where the method of \citet{lyman14} 
does not apply, we estimate 
(pseudo)-bolometric corrections using the late-time SN 
spectra. First we calibrate the two Keck~I 
spectra using the $r$-band photometry. We 
then integrate the entire flux between 3000 
and 9800 \AA\ in the spectra. This flux, 
transformed into magnitudes and subtracted by 
the $r$-band magnitude from the photometry at 
the same epoch, gives a (pseudo) bolometric 
correction that can be applied to the $r$ 
band to compute a (quasi) bolometric light 
curve. We use the spectrum at 360~d to 
compute the pseudo-bolometric correction  
between 300~and 500~d, and the spectrum 
obtained at 650~d to compute the pseudo-bolometric correction at later epochs. We 
then apply these corrections to the $r$-band 
light curve to obtain the bolometric light 
curve plotted in Fig.~\ref{bolofit}. We note 
that with our method our pseudo-bolometric 
light curve at late epochs does not include 
the ultraviolet and near-infrared flux, and 
can therefore be regarded as a lower limit of 
the actual bolometric luminosity. Despite 
this fact, it is still remarkably brighter 
than the expected $^{56}$Co luminosity at 
late epochs (see Sect.~\ref{sec:model}).

\section{Modeling}
\label{sec:model}

\begin{figure*}
\includegraphics[width=18cm]{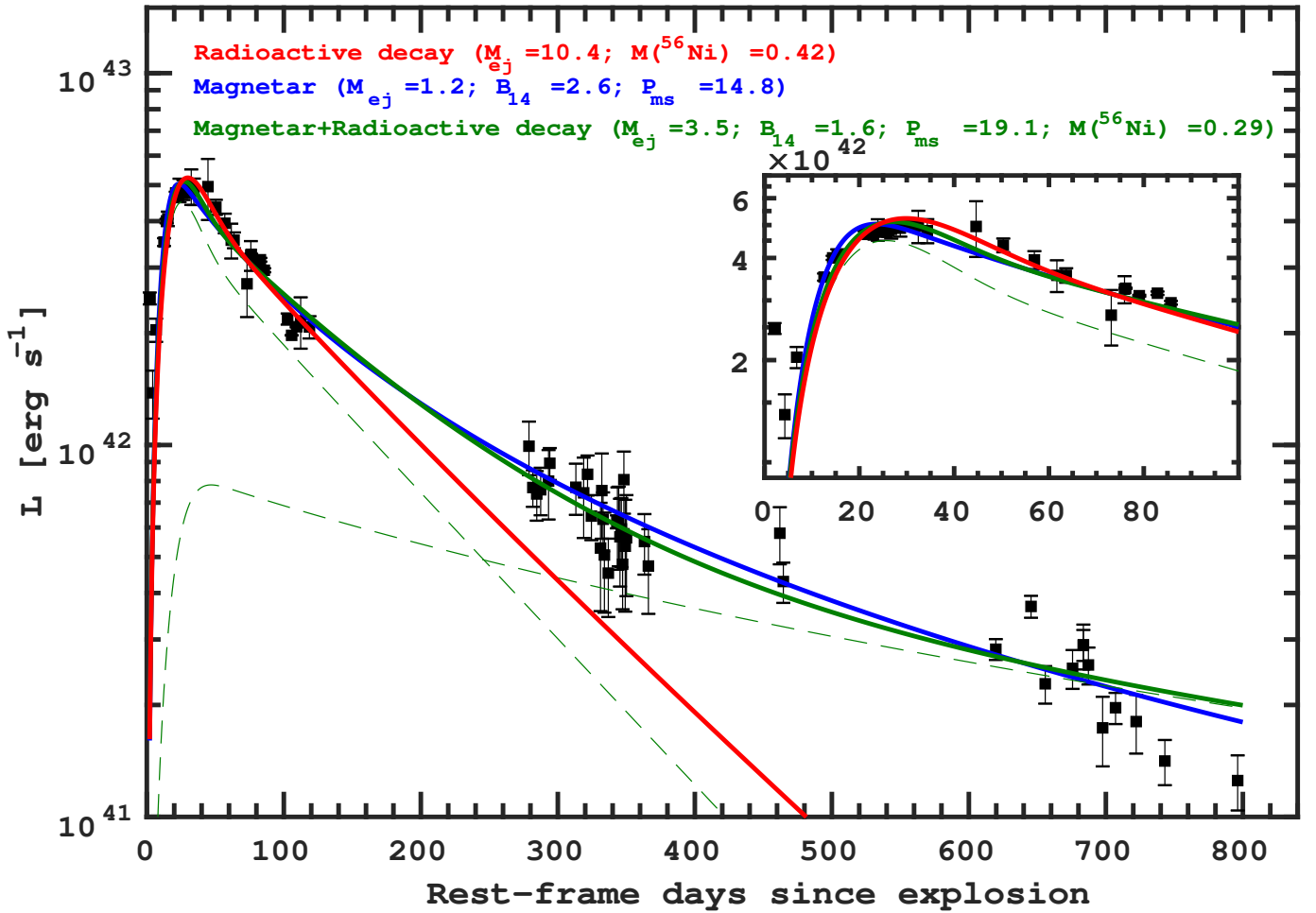}
\caption{\label{bolofit}Bolometric light curve of iPTF15dtg from explosion until 800~d (rest-frame). For the first 150~d we adopt the bolometric light curve that was computed in T16. After that, we make use of our spectra to compute late-time bolometric corrections to the $r$-band light curve and hence obtain the luminosity at late epochs. We fit the bolometric light curve with three models: a purely radioactive model (red), a magnetar model (blue) and a hybrid model (green). The best fit parameters are reported in the corresponding colors. The two components of the hybrid model are shown by green dashed lines (at early epochs the radioactivity dominates, at late epochs the magnetar is prominent). The inset shows a zoomed view of the fits around the light curve peak.}
\end{figure*}

We attempt to model the bolometric light curve with three different models. The three models are shown in Fig.~\ref{bolofit}. 

 First we confirm that radioactive decay alone cannot explain the late-time
 light curve of iPTF15dtg. Even though it provides a good fit to the
 early light curve (as also shown in T16), it is not possible to reproduce the excess at the late
 epochs with a simple \citep{arnett82} model including 0.42 $M_{\odot}$ of $^{56}$Ni and a large ejecta mass of 10.4 $M_{\odot}$  (red line in Fig.~\ref{bolofit}). Here, following T16 and the measurements of the \ion{Fe}{ii} line P-Cygni absorption velocity, we assumed the ejecta to expand with a bulk velocity of 6000 km~s$^{-1}$ and to have uniform density. These assumptions imply a kinetic energy of  2.2$\times$10$^{51}$~erg for the radioactive model.

An additional source of power is needed to fit the early and the late-time light curves simultaneously. We test if a magnetar can fit both
phases (blue line in Fig.~\ref{bolofit}). Indeed, using the models presented in \citet{inserra13} and
derived from \citet{kasen10}, we obtain a good fit with an ejecta mass
of 1.2~$M_\odot$, magnetic field intensity
B$= 2.6 \times$10$^{14}$~G, and an initial spin period of 14.8~ms. However, the small ejecta mass required to fit the late epochs does not perfectly reproduce the early peak, which is rather broad. This model assumes again an ejecta bulk velocity of 6000 km~s$^{-1}$ and uniform density, implying a low kinetic energy of 0.2$\times$10$^{51}$~erg).

The  model with the magnetar alone also produces an earlier peak than what is observed (see inset of Fig.~\ref{bolofit}). Furthermore, we know that iPTF15dtg is somewhat similar to
other SE SNe. It is not super-luminous and displays a perfectly normal spectral evolution. It is therefore reasonable to assume that it
ejected some amount of
$^{56}$Ni, like other SE~SNe. 
This $^{56}$Ni will not strongly influence the late-time light curve, where another source, like a magnetar, will dominate.
Therefore, we try a hybrid model where the input powering is both a magnetar and radioactive $^{56}$Ni (green lines in Fig.~\ref{bolofit}). In this case, we obtain a better fit at earlier epochs (a longer rise time, see solid green light in the inset of Fig.~\ref{bolofit}) where the $^{56}$Ni dominates, 
and a good fit at late epochs where a magnetar is prominent.
The best-fit ejecta mass turns out to be about 3.5 $M_\odot$, the $^{56}$Ni mass is 0.29~$M_\odot$, the spin period 19.1~ms, and the magnetic field 
B$= 1.6 \times 10 ^{14}$~G. Given the same assumptions on the ejecta velocity as in the other models, the kinetic energy turns out to be 0.8$\times$10$^{51}$~erg. 
These are physically reasonable values, which describe iPTF15dtg as a normal SE~SN where a magnetar modifies the peak and dominates the emission at late epochs, making this SN extraordinarily bright more than two years after explosion. 

The modeling presented in this section does not include any possible gamma-ray and spin-down energy leakage, as presented in \citet{wang15}. This could play an important role in the late-time light curve. We test this by running a few models (both magnetar and magnetar$+$radioactivity) allowing for such energy leakage. We consider gamma-ray opacities between 0.01 and 0.2 
(0.01, 0.03, 0.1, 0.2)~cm$^2$~g$^{-1}$, as suggested in \citet{wang15}. The hybrid model provides a good fit for $\kappa_{\gamma}$ = 0.2~cm$^2$~g$^{-1}$, and a moderately good fit for $\kappa_{\gamma}$ = 0.1~cm$^2$~g$^{-1}$. For lower gamma-ray opacities it does not provide a good fit. The simple magnetar model provides a moderately good fit for $\kappa_{\gamma}$ = 0.1~cm$^2$~g$^{-1}$ and $\kappa_{\gamma}$ = 0.03~cm$^2$~g$^{-1}$. All these fits are however poorer than those obtained assuming full trapping. If we consider the best model among those including energy leakage, that is, the hybrid model with $\kappa_{\gamma}$ = 0.2~cm$^2$~g$^{-1}$, this provides an ejecta mass of 3.4~M$_{\odot}$, similar to the case of the full-trapping hybrid model, but with a much lower (more than two orders of magnitude) magnetic-field intensity and spin period (one order of magnitude). At epochs later than $\sim$700~d the bolometric light curve starts to decline faster as compared to the previous $\sim$500~d, and the hybrid model that allows for the energy leakage fits this phase
better than the models that do not allow for the leakage. 
 If we consider the best magnetar model ($\kappa_{\gamma}$ = 0.1~cm$^2$~g$^{-1}$), again we find a low ejecta mass (0.5~M$_{\odot}$), relatively similar to the best magnetar model with full trapping, but a much lower $B$ and spin period (two and one order of magnitude, respectively). We note that the required gamma-ray opacity $\kappa_{\gamma}$ to get a good late-phase magnetar fit (0.2/0.1~cm$^{2}$~g$^{-1}$) is relatively high compared with those estimated for the SLSN magnetar fits.

Also, T16 examined the possibility that the very early light-curve excess shown by iPTF15dtg could arise from a magnetar-powered shock breakout \citep{kasen16}. However, for the parameters derived from both the magnetar model and the hybrid model, the maximum luminosity of this first peak would be one order of magnitude larger than what is observed, and the same is true for the timescale of the breakout. This is true also for the models where we do not assume full trapping.

\section{Discussion}
\label{sec:discussion}

\subsection{Powering mechanism}
To conclude on the powering mechanism of iPTF15dtg, this cannot be only due to the regular radioactive decay powering as in other SE SNe, given its extraordinary slow late-time decline. If the bolometric light curve between 300 and 400~d were entirely powered by radioactivity, a $^{56}$Ni mass of 0.8 M$_{\odot}$ would be required. However, this amount of radioactive material would also imply a peak bolometric light curve of 10$^{43}$~erg~s$^{-1}$, which is a factor of two brighter than what we observe. We also note that other isotopes such as $^{57}$Co and $^{44}$Ti do not contribute significantly to the light curve until epochs later than our observations \citep[see][and the case of SN~1987A]{fransson93}, so they cannot account for the excess of iPTF15dtg. 

A magnetar alone might power the light curve of iPTF15dtg. Our magnetar model fits reasonably well the evolution across the entire 800-d period during which we observed iPTF15dtg. However, the fit at early epochs is not perfect, as the peak epoch of iPTF15dtg occurs later than that of our magnetar model. 
Furthermore, we must consider that SN iPTF15dtg 
in many respects is similar to a normal SE SNe, and thus we expect to have some radioactive material also contributing to the luminosity. Therefore, a hybrid model with radioactive material dominating at early epochs and a magnetar dominating the nebular phase is our favorite scenario and gives the best fit to the bolometric light curve. 
We note that when we consider the possibility of spin-down energy leakage in our models instead of full trapping, we generally obtain lower-quality fits, but we can however reproduce the light curve obtaining similar ejecta masses to the models that assume full trapping. However, these models are characterized by a much lower magnetic field intensity and spin period.

Based on the spectral and light-curve comparison to the interacting SE SN~2010mb, we consider it less likely that circumstellar interaction plays a major role as a source of luminosity for iPTF15dtg, even though we cannot completely exclude its presence. We argue that the spectra of iPTF15dtg are too similar to those of other regular SE~SNe to hide a CSM interaction responsible for an excess in luminosity of several magnitudes. However there are a few arguments that might be in favor of the CSM-interaction mechanism. 
First, both magnetar and radioactive powering are considered as smooth processes that produce a smoothly declining luminosity. However, the light curve of iPTF15dtg shows some variability and a late-time break that might be more naturally explained by CSM interaction. However, it is not excluded that a magnetar-powered SN can
produce fluctuations in the light curve, as observed in some SLSNe-I \citep[e.g.,][]{nicholl16}.
We also noted that the spectra of iPTF15dtg at late times have a blue excess that is slightly more prominent than in other SE SNe, even though it is not as pronounced as in SN~2010mb. This excess looks quite similar to the late emission from SNe IIn (except for the H lines), where the blue excess is dominated by \ion{Fe}{ii}
emission from CSM interaction. However, the light-curve
excess is too strong to be explained by this blue excess. 
The late-time luminosity may also be explained in terms of CSM interaction, adopting for example a model from \citet{chatzopoulos13} with the collision between the ejecta and a CSM shell of 0.1~$M_{\odot}$ occurring 130~d post-explosion. However, such a scenario is difficult to reconcile with the spectral observations as previously explained.

Given the long-lasting emission of iPTF15dtg we might also consider more exotic powering mechanisms, such as fallback-accretion powering. In the case of iPTF14hls, an extraordinary spectroscopically normal SN~II with a very long-lasting emission, the late-time bolometric light curve was compared to a power law with index $-5/3$ \citep{sollerman18}, which is expected for a fallback-accretion-powering scenario. If we compare the light curve of iPTF15dtg with this expected trend, we find that the emission of iPTF15dtg at epochs later than 250~d are overall in agreement with such a power law. However,
we notice that in the late-time spectra of iPTF15dtg we detect strong [\ion{O}{i}] emission. Oxygen, which sits in the inner layers of the ejecta, should be among the first elements to fall back into the central black hole, and therefore should not leave a strong signature in the spectra.
Another possibility is that the late-time emission of iPTF15dtg is powered by a light echo \citep[see e.g.,][]{chevalier86}; however without near-infrared observations it is hard to assess this scenario.

\subsection{Progenitor implications}

T16 observed iPTF15dtg up to around 150~d, and favor a radioactive model for this SN, with a large ejecta mass of 10.3 $M_{\odot}$ and a $^{56}$Ni mass of 0.42 $M_{\odot}$.
This implies a massive, possibly single progenitor star of at least 35 $M_{\odot}$, something very unusual among the SE SNe. However, here we show that the late-time bolometric light curve of iPTF15dtg displays - for the first time for a normal SE SN - a decline rate much slower than that of $^{56}$Co, implying a different or additional mechanism powering the luminosity. Our hybrid model, where the contribution of a magnetar is also included and which provides the best fit to the data, implies a lower ejecta mass for iPTF15dtg of about 3.6 $M_{\odot}$. This is in line with what has been presented in the literature for regular SE~SNe \citep[see e.g.,][]{taddia15,lyman16,taddia18} and might imply a lower initial mass for the progenitor star of iPTF15dtg, and therefore a likely origin in a binary system given that this SN was stripped of its H and He envelopes. 

There is a caveat to this result. If the magnetar spin-down energy is not fully trapped, which we are assuming in our models, then a larger ejecta mass would be required to sustain such a slow luminosity decline, and a (single) massive star progenitor scenario for SN iPTF15dtg might still be allowed. However, if we include in our model the leakage of hard emission as presented by \citet{wang15} for SLSNe powered by magnetars, we would not be able to fit both the peak and tail of the bolometric light curve of iPTF15dtg simultaneously. The reason is that the leakage would push the late-time luminosity of our model towards lower values than those observed.  Even for a large ejecta mass of 10~$M_{\odot}$, the luminosity at 800~d would be reduced by a factor of four, preventing a good fit of the light-curve tail.  
However, it is not clear that the leakage of energy from a magnetar occurs as leakage of hard emission, as assumed by \citet{wang15}. In fact the assumption of full energy trapping might not apply after 700~d, where the light curve begins to drop faster.

\section{Conclusion}
\label{sec:concl}

We observed supernova iPTF15dtg until $\sim$840~d post-explosion. The spectra are similar to those of regular SE~SNe. On the contrary, the late-time light curve declines very slowly compared to $^{56}$Co decay, which is unprecedented among regular SE SNe. The bolometric light curve is well reproduced by a model including both radioactivity (dominating at early epochs) and a magnetar (powering the late emission). This model allows us to revise the ejecta mass of iPTF15dtg to a lower value compared to what was discussed in T16, implying a lower initial mass and thus a possible binary scenario for the progenitor star of iPTF15dtg. 

\begin{acknowledgements}
FT and JS gratefully acknowledge support from the Knut and Alice Wallenberg Foundation. JS acknowledges the support of Vetenskapsr\aa det through VR grants 
2012-2265 and 2017-03699. 
This work is partly based on observations made with the Nordic Optical Telescope, operated by the Nordic Optical Telescope Scientific Association at the Observatorio del Roque de los Muchachos, La Palma, Spain, of the Instituto de Astrofisica de Canarias.
The data presented here were obtained in part with ALFOSC, which is provided by the Instituto de Astrofisica de Andalucia (IAA) under a joint agreement with the University of Copenhagen and NOTSA.
This work is partly based on observations made with DOLoRes@TNG. 
Based  on  observations  obtained  with  the  Samuel  Oschin
Telescope 48-inch and the 60-inch Telescope at the Palomar
Observatory  as  part  of  the  intermediate  Palomar  Transient
Factory (iPTF) project,  a scientific collaboration among the
California Institute of Technology, Los Alamos National Laboratory, the University of Wisconsin, Milwaukee, the Oskar
Klein Center, the Weizmann Institute of Science, the TANGO
Program of the University System of Taiwan, and the Kavli
Institute  for  the  Physics  and  Mathematics  of  the  Universe.

\end{acknowledgements}

\bibliographystyle{aa}

\onecolumn

\begin{deluxetable}{|ccc|ccc|ccc|}
\tabletypesize{\scriptsize}
\tablewidth{0pt}
\tablecaption{Late-time optical photometry of iPTF15dtg.\label{tab:phot}}
\tablehead{
\colhead{JD}&
\colhead{$g$}&
\colhead{$err_g$}&
\colhead{JD}&
\colhead{$r$}&
\colhead{$err_r$}&
\colhead{JD}&
\colhead{$i$}&
\colhead{$err_i$}\\
\colhead{(days)}&
\colhead{(mag)}&
\colhead{(mag)}&
\colhead{(days)}&
\colhead{(mag)}&
\colhead{(mag)}&
\colhead{(days)}&
\colhead{(mag)}&
\colhead{(mag)}}
\startdata
\multicolumn{9}{c}{{\bf P48}}\\
2457642.97 &  20.95 &  0.24 &  2457627.00 & 20.07 & 0.18 &  &  &   \\
2457656.99 &  20.94 &  0.30 &  2457629.99 & 20.35 & 0.12 &  &  &   \\
2457662.99 &  20.72 &  0.16 &  2457632.98 & 20.40 & 0.17 &  &  &   \\
2457665.98 &  20.74 &  0.07 &  2457635.99 & 20.37 & 0.16 &  &  &   \\
2457668.97 &  20.70 &  0.29 &  2457642.02 & 20.31 & 0.23 &  &  &   \\
2457672.03 &  20.44 &  0.29 &  2457643.01 & 20.19 & 0.11 &  &  &   \\
2457674.98 &  20.73 &  0.21 &  2457663.02 & 20.35 & 0.17 &  &  &   \\
2457687.90 &  21.23 &  0.34 &  2457669.02 & 20.39 & 0.26 &  &  &   \\
2457693.89 &  20.78 &  0.18 &  2457671.99 & 20.26 & 0.13 &  &  &   \\
2457694.91 &  21.20 &  0.36 &  2457674.96 & 20.54 & 0.15 &  &  &   \\
2457695.90 &  20.80 &  0.29 &  2457681.95 & 20.76 & 0.35 &  &  &   \\
2457696.88 &  21.01 &  0.20 &  2457682.95 & 20.37 & 0.28 &  &  &   \\
2457697.91 &  20.89 &  0.19 &  2457683.92 & 20.57 & 0.20 &  &  &   \\
2457698.88 &  20.81 &  0.16 &  2457684.95 & 20.81 & 0.33 &  &  &   \\
2457699.87 &  21.04 &  0.29 &  2457687.94 & 20.93 & 0.26 &  &  &   \\
2457700.90 &  21.03 &  0.30 &  2457694.95 & 20.59 & 0.27 &  &  &   \\
           &        &       &  2457695.94 & 20.58 & 0.26 &  &  &   \\
           &        &       &  2457696.92 & 20.68 & 0.29 &  &  &   \\
           &        &       &  2457697.95 & 20.63 & 0.23 &  &  &   \\
           &        &       &  2457698.93 & 20.87 & 0.26 &  &  &   \\
           &        &       &  2457699.91 & 20.30 & 0.21 &  &  &   \\
           &        &       &  2457700.94 & 20.75 & 0.36 &  &  &   \\
           &        &       &  2457701.91 & 20.69 & 0.33 &  &  &   \\
           &        &       &  2457715.74 & 20.72 & 0.20 &  &  &   \\
           &        &       &  2457718.73 & 20.88 & 0.28 &  &  &   \\

\hline
\multicolumn{9}{c}{{\bf P60}}\\
 &  & & 2457819.69 &  20.66  &   0.19 & & & \\
\hline
\multicolumn{9}{c}{{\bf NOT}}\\
    2457694.55 & 21.26 & 0.08 & 2457694.55 & 20.58 & 0.04 & 2457694.55 & 20.73 & 0.05\\
    2457822.37 & 21.42 & 0.20 & 2457822.38 & 20.98 & 0.13 & 2457822.38 & 21.22 & 0.16\\
    2457985.58 & 22.02 & 0.19 & 2457985.59 & 21.60 & 0.07 & 2457985.59 & 21.97 & 0.11\\
    2458044.54 & 21.93 & 0.17 & 2458044.55 & 21.72 & 0.13 & 2458044.56 & 22.12 & 0.23\\
    2458056.71 & 22.02 & 0.16 & 2458056.72 & 21.70 & 0.12 & 2458056.73 & 22.33 & 0.21\\
               &              & &2458067.64 & 22.12 & 0.23 &            &       &      \\
               &              & &2458093.42 & 22.08 & 0.20 &            &       &      \\   
\hline
\multicolumn{9}{c}{{\bf TNG}}\\
      
         2458012.63 & 21.54 & 0.07 & 2458012.65 & 21.31 & 0.07 & 2458012.64 & 21.48 & 0.07 \\
     2458023.56 & 21.54 & 0.08 & 2458023.57 & 21.83 & 0.13 & 2458023.58 & 21.25 & 0.05 \\            
     2458077.55 & 22.26 & 0.18 & 2458077.57 & 21.99 & 0.10 & 2458115.43 & 22.56 & 0.19 \\
     2458115.42 & 22.51 & 0.21 & 2458115.44 & 22.35 & 0.15 & 2458171.34 & 22.87 & 0.21 \\
     2458171.35 & 22.84 & 0.18 & 2458171.36 & 22.48 & 0.18 &            &       &      \\
\hline
\multicolumn{9}{c}{{\bf P200}}\\
2458052.84 & 21.83 & 0.29 &  2458052.85 &  21.57 &  0.15 & 2458052.85 & 22.20 & 0.13 \\
           &       &      &  2458052.85 &  21.56 &  0.10 &            &       &      \\
  \enddata
\end{deluxetable}

\begin{deluxetable}{cccccc}
\tabletypesize{\scriptsize}
\tablewidth{0pt}
\tablecaption{Late-time optical spectroscopy of iPTF15dtg\label{tab:spectra}}
\tablehead{
\colhead{Date (UT)}&
\colhead{JD-2,457,000}&
\colhead{Phase\tablenotemark{a}}&
\colhead{Telescope}&
\colhead{Instrument}&
\colhead{Range}\\
\colhead{}&
\colhead{(days)}&
\colhead{(days)}&
\colhead{}&
\colhead{}&
\colhead{(\AA)}}
\startdata
31 Oct 2016    &   693.02    &  +359.6   & Keck1 &   LRIS     & 3080$-$10302    \\
07 Dec 2016    &   730.53    &  +397.1   & NOT   &   ALFOSC   & 3801$-$9711     \\
17 Aug 2017    &   982.97    &  +649.5   & Keck1 &   LRIS     & 3071$-$10268    \\
16 Sep 2017    &   1012.74   &  +679.3   & TNG   &   DOLORES  & 3801$-$9719    \\
26 Sep 2017    &   1023.71   &  +690.3   & TNG   &   DOLORES  &  3501$-$8341   \\
13 Nov 2017    &   1070.83   &  +737.4   & Keck2 &   DEIMOS   & 4403$-$9905      \\
\enddata
\tablenotetext{a}{Days from explosion date in the observer frame.}
\end{deluxetable}

\end{document}